%% file: main.tex
\documentclass[aps, prb, 10pt, twocolumn, nofootinbib, superscriptaddress, longbibliography]{revtex4-2}

\input{preamble}

\begin{document}

\title{Approaching the Kosterlitz-Thouless transition for the \\ classical XY model with tensor networks}

\author{Laurens Vanderstraeten}
\email{laurens.vanderstraeten@ugent.be}
\affiliation{Department of Physics and Astronomy, University of Ghent, Krijgslaan 281, 9000 Gent, Belgium}

\author{Bram Vanhecke}
\affiliation{Department of Physics and Astronomy, University of Ghent, Krijgslaan 281, 9000 Gent, Belgium}

\author{Andreas M. L\"{a}uchli}
\affiliation{Institute for Theoretical Physics, University of Innsbruck, 6020 Innsbruck, Austria}

\author{Frank Verstraete}
\affiliation{Department of Physics and Astronomy, University of Ghent, Krijgslaan 281, 9000 Gent, Belgium}

\begin{abstract}
We apply variational tensor-network methods for simulating the Kosterlitz-Thouless phase transition in the classical two-dimensional XY model. In particular, using uniform matrix product states (MPS) with non-abelian $\mathrm{O}(2)$ symmetry, we compute the universal drop in the spin stiffness at the critical point. In the critical low-temperature regime, we focus on the MPS entanglement spectrum to characterize the Luttinger-liquid phase. In the high-temperature phase, we confirm the exponential divergence of the correlation length and estimate the critical temperature with high precision. Our MPS approach can be used to study generic two-dimensional phase transitions with continuous symmetries.
\end{abstract}

\maketitle

\section{Introduction}

In contemporary theoretical physics the interplay between symmetry and dimensionality is widely appreciated for giving rise to fascinating physical phenomena. Historically, one of the crucial results in establishing this viewpoint was the phase transition in the two-dimensional classical XY model. This model is introduced by placing continuous angles $\{\theta_i\}$ on a square lattice and characterizing their interactions by the hamiltonian
\begin{equation*}
H = - \sum_{\braket{ij}} \cos\left(\theta_i-\theta_j\right) - h \sum_i \cos(\theta_i),
\end{equation*}
where $\braket{ij}$ labels all nearest-neighbour pairs and we introduce an external magnetic field $h$ for further reference.
\par The first ingredient for understanding the phase diagram of the XY model (without magnetic field) is the Mermin-Wagner theorem \cite{Mermin1966, Hohenberg1967}: No conventional long-range order can exist at any finite temperature in this model, because of the proliferation of spin-wave excitations in two dimensions. Still, one expects a phase transition in this system. At low temperatures, the system can be described by a simple continuum field theory with algebraically decaying correlation functions. A high-temperature expansion, however, suggests an exponential decay of correlations in this system at sufficiently high temperature. As Berezinskii \cite{Berezinskii1972} and Kosterlitz and Thouless \cite{Kosterlitz1973} (BKT) have showed, the phase transition between the low- and high-temperature regime is driven by the unbinding of vortices and is, therefore, of a topological nature. The transition is captured by a renormalization-group (RG) analysis \cite{Kosterlitz1974, Jose1977}, where in the critical phase the long-wavelength properties are described by a free bosonic field theory with continuously varying critical exponents, until vortices become relevant and drive the system into a gapped phase at a critical temperature $T_c$. The quantity that characterizes the transition (in the absence of a local order parameter) is the spin stiffness $\rho(T)$, which discontinuously drops \cite{Nelson1977} from a finite value in the critical phase to zero in the gapped phase; from the renormalization analysis, the size of the drop is known to be given by
\begin{equation} \label{eq:stiff}
\lim_{T\to T_c^-} \rho(T) = \frac{2T_c}{\pi} ,
\end{equation}
but the value of $T_c$ is not known exactly. When approaching the critical point from the gapped side, the BKT transition is characterized by an exponential divergence of the correlation length \cite{Kosterlitz1974, Minnhagen1987},
\begin{equation} \label{eq:corr}
\xi(T) \propto \exp\left( \frac{b}{\sqrt{T-T_c}} \right), \qquad T \to T_c^+,
\end{equation}
with $b$ a non-universal parameter.
\par The BKT phase transition in the XY model has been a notoriously hard case for numerical simulations, because of the exponentially diverging correlation length and the ensuing logarithmic finite-size corrections around the phase transition. Nonetheless, shortly after the theoretical work, Monte-Carlo simulations \cite{Miyashita1978, Tobochnik1979} provided considerable evidence for the correctness of the theory and rough estimates for the transition temperature. In recent years, these estimates were significantly improved \cite{Hasenbusch2005, Komura2012}, obtaining a value around $T_c\approx0.8929$, in agreement with high-temperature expansions \cite{Arisue2009}. Still, these results depend heavily on assumptions about the logarithmic finite-size corrections and an improved extrapolation \cite{Hsieh2013} places the transition temperature at a higher value of $T_c\approx0.8935$. Crucially, the best estimates use the universal value for the spin stiffness at the phase transition for pinpointing the critical point.
\par Tensor networks \cite{Verstraete2008, Orus2013} provide an original framework for capturing the symmetry-dimensionality interplay, both from the theoretical and the numerical side. Although orginally devised for capturing the entanglement in strongly-correlated quantum lattice systems, tensor networks are increasingly being applied to problems in statistical mechanics. Since the framework is entirely different from traditional approaches such as Monte-Carlo sampling, it can shed a new, entanglement-based, light on statistical-mechanics problems. Indeed, tensor networks encode all physical properties of a given system into local tensors, and they allow to understand the relation between physical symmetries of the local degrees of freedom and the global properties of the system in a transparent way. One particular example is the classification of symmetry-protected topological phases in one-dimensional quantum systems, which is brought back to the symmetry properties of the local tensors that make up a matrix product state (MPS) \cite{Pollmann2010, Chen2011}. This was extended to two dimensions, where the topological order of a wavefunction can be related to the symmetry properties of the local tensor in a projected entangled-pair state (PEPS) \cite{Schuch2010}. On the numerical side, physical symmetries can be explicitly incorporated in tensor-network algorithms and lead to an improved efficiency and performance \cite{Mcculloch2002, Singh2010, Weichselbaum2012, ZaunerStauber2018}.
\par The encoding of symmetries in tensor networks is performed most elegantly when working directly in the thermodynamic limit, because one can purely focus on the symmetry properties of the bulk tensors without bothering about what happens at the boundaries of the system. Yet, applying uniform tensor networks with an explicit encoding of the physical symmetries seem to break down when considering critical phases. For example, in a critical phase a uniform MPS \cite{Draxler2013} typically favours an artificial breaking of a continuous symmetry, where the associated order parameter decreases very slowly as the bond dimension is increased. The apparent reason for this artificial symmetry breaking is that MPS have a built-in limitation for the amount of entanglement in the state, which makes it energetically favourable to break a continuous symmetry. This seems to imply that uniform tensor networks fail to capture the essential properties of critical phases with a continuous symmetry.
\par In this paper, we explore this question in more detail by investigating the precise sense in which uniform MPS capture critical phases with a continuous symmetry. As explained, the XY model serves as the paradigmatic example of a system where the absence of symmetry breaking leads to a critical phase, and, therefore, we take the XY model as our test case. In contrast to earlier tensor-network approaches for the XY model \cite{Yu2014}, we use uniform MPS methods for transfer matrices \cite{Haegeman2016b, Fishman2018, Vanderstraeten2019} as a means for characterizing the BKT phase transition. In the first two sections we explain the duality transformation \cite{Jose1977} that allows us to define a row-to-row transfer matrix, approximate its fixed point as a uniform MPS and to compute local observables. In the next section, we focus on the spin stiffness as the characteristic quantity in the BKT phase transition. Afterwards, we use the Luttinger-liquid formalism to characterize the critical phase. Finally, in the last section, we focus on the gapped phase and locate the critical temperature with high precision.

\section{MPS for the XY transfer matrix}

\subsection{Partition function}

We start by writing down the partition function for the XY model as a tensor network. The partition function at a given inverse temperature $\beta=1/T$ is given by
\begin{equation*}
Z =   \prod_i \int \frac{\d\theta_i}{2\pi} \prod_{\braket{ij}} \e^{\beta\cos(\theta_i-\theta_j)} \prod_i \e^{\beta h \cos(\theta_i)} .
\end{equation*}
In order to arrive at a tensor-network representation, we introduce a duality transformation \cite{Jose1977} that maps the above partition function to a representation in terms of bosonic degrees of freedom on the links. Such a map is obtained by introducing the following decomposition on every link in the lattice
\begin{equation*}
\e^{x\cos(\theta_i-\theta_j)} = \lim_{N\rightarrow\infty} \sum_{n=-N}^N I_n(x) \e^{in(\theta_i-\theta_j)},
\end{equation*}
where $I_{n}(x)$ are the modified Bessel functions of the first kind. Then, by integrating over all the $\theta$'s, the partition function is transformed into
\begin{equation*}
Z = \lim_{N\to\infty} \prod_{l\in\mathcal{L}} \left( \sum_{n_l=-N}^N I_{n_l} (\beta) \right)  \prod_{s} F_{n_{s,1},n_{s,2}}^{n_{s,3},n_{s,4}}
\end{equation*}
where $F$ is a four-index tensor
\begin{equation*}
F_{n_1,n_2}^{n_3,n_4} = \int \frac{\d\theta}{2\pi} \; \e^{\beta h \cos\theta} \e^{i\theta(n_{1}+n_{2}-n_{3}-n_{4})}.
\end{equation*}
The first product runs over all the links in the lattice, and $s$ labels all the sites in the lattice. We can now represent this partition function as a network of tensors, 
\begin{equation*}
Z = \diagram{intro}{1},
\end{equation*}
where every tensor $O$ is given by
\begin{equation*}
\diagram{intro}{2} =  \left( \prod_{i=1}^4I_{n_i}(\beta) \right)^{1/2}   F_{n_1,n_2}^{n_3,n_4}
\end{equation*}
and the virtual legs $n_i$ have infinite dimension. In practice, however, it will be possible to truncate these indices without loss of accuracy. We have introduced arrows on the legs to indicate the signs in which the $n_i$'s appear in the $F$ tensor above.
\par The fundamental object in this representation of the partition function is the row-to-row transfer matrix $T(\beta,h)$
\begin{equation*}
T(\beta,h) = \diagram{intro}{3} \;,
\end{equation*}
which is an operator acting on an infinite chain of bosonic degrees of freedom. The value for the partition function and, therefore, the free energy of the model, is determined by the leading eigenvalue $\Lambda$ of the transfer matrix. Indeed, this leading eigenvalue is expected to scale as the number of sites per row, i.e. $\Lambda \sim \lambda^{N_x}$, such that the free energy per site is
\begin{align*}
f(\beta,h) &=  \lim_{N_xN_y\to\infty} \left( - \frac{1}{\beta} \frac{1}{N_xN_y} \log Z \right) \\
&=  \lim_{N_xN_y\to\infty} \left( -  \frac{1}{\beta}  \frac{1}{N_xN_y} \log \left( T(\beta,h)^{N_y} \right) \right) \\
&= - \frac{1}{\beta} \log \lambda(\beta,h).
\end{align*}
\par
\begin{figure} 
\begin{center}
\includegraphics[width=0.95\columnwidth]{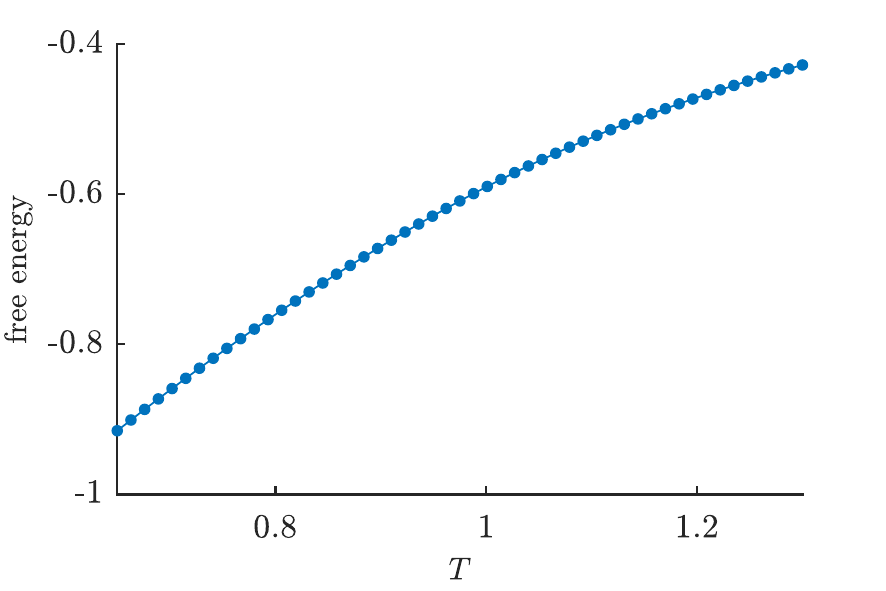}
	\caption{Free energy per site of the XY model ($h=0$) as obtained from variational MPS simulations with bond dimension $D=30$ and without using any symmetries on the virtual legs.}
    \label{fig:free}
\end{center}
\end{figure}

 The eigenvector corresponding to the leading eigenvalue, refered to as the fixed point of the transfer matrix,
\begin{equation*}
T(\beta,h) \ket{\Psi_{\beta,h}} = \Lambda \ket{\Psi_{\beta,h}},
\end{equation*}
will be of crucial importance in all computations. In a number of applications, it has been shown that fixed points of transfer matrices can be approximated accurately using the variational class of matrix product states (MPS) \cite{Haegeman2016b}. For translation-invariant transfer matrices, we can describe the fixed point as a uniform MPS described by a single tensor $A$,
\begin{equation*}
\ket{\Psi(A)} = \diagram{fixedpoint}{1}.
\end{equation*}
The tensor $A$ has virtual legs of dimension $D$, which we call the MPS bond dimension, and by repeating the tensor on every site and contracting over the virtual legs, we obtain a translation-invariant state. For simplicity, we take the virtual legs of the MPS to have no arrows. In terms of the fixed-point MPS, the fixed-point eigenvalue equation is rephrased as
\begin{multline*}
\diagram{fixedpoint}{2} \propto \\ \diagram{fixedpoint}{1}.
\end{multline*}
We aim at finding a tensor for which this eigenvalue problem is obeyed in an optimal way. Since the transfer matrix is hermitian, we can use the variational principle for establishing an optimization problem for the tensor $A$:
\begin{equation*}
\max_A \frac{\bra{\Psi(\bar{A})} T(\beta,h) \ket{\Psi(A)} }{\braket{\Psi(\bar{A}) | \Psi(A)}}.
\end{equation*}
This optimization problem can be efficiently solved using tangent-space methods for uniform MPS \cite{Haegeman2016b, Vanderstraeten2019}; in particular, we use the vumps algorithm \cite{Fishman2018} for finding the optimal MPS tensor.  The eigenvalue, and therefore the free energy, is then obtained as the contraction of an infinite channel of $O$ tensors sandwiched between the fixed-point MPS and its conjugate,
\begin{align*}
\Lambda &= \bra{\Psi(\bar{A})} T(\beta,h) \ket{\Psi(A)} \\
&= \diagram{fixedpoint}{3},
\end{align*}
or we find $\lambda$ as the leading eigenvalue (spectral radius $\rho$) of the channel operator
\begin{equation*}
\lambda = \rho\left( \diagram{fixedpoint}{4} \right).
\end{equation*}
Here, we have assumed that the MPS itself is normalized as
\begin{equation*}
\rho\left( \diagram{fixedpoint}{5} \right) = 1.
\end{equation*}
In Fig.~\ref{fig:free} we have plotted the result for the free energy per site as obtained by a variational MPS simulation of the XY model in a certain temperature range and without magnetic field, where we use MPS with a bond dimension of $D=30$. It is clear that the free energy shows no signs of a phase transition as it is perfectly smooth everywhere. Note that, whereas the free energy cannot be directly computed using Monte-Carlo sampling, it appears as the fundamental quantity in the variational MPS setup.

\subsection{Symmetries in the MPS representation}
\label{sec:symmetries}

The case without magnetic field ($h=0$) is of particular importance, because the model is invariant under a global transformation $\theta_i\to\theta_i+\alpha$. This $\mathrm{U}(1)$ invariance is reflected in the tensor-network representation as a symmetry of the transfer matrix. Indeed, if $h=0$ we have
\begin{align*}
\left. F_{n_1,n_2}^{n_3,n_4} \right|_{h=0} &= \int \frac{\d\theta}{2\pi} \; \e^{i\theta(n_{1}+n_{2}-n_{3}-n_{4})} \\
&= \delta_{n_1+n_2}^{n_3+n_4}.
\end{align*}
such that the tensor $O$ has conservation of $\mathrm{U}(1)$ charges. We can introduce the operator $Q$
\begin{equation} \label{eq:Qtensor}
\diagram{intro}{4} = \delta_{n_1}^{n_2} n_1,
\end{equation} 
which counts the charge on a given leg in the tensor network; this operator is the generator of the $\mathrm{U}(1)$ symmetry of the transfer matrix. Indeed, from the conservation property for the tensor $O$ the transfer matrix clearly commutes with the symmetry operation
\begin{align*}
U(\theta) = \e^{i \theta \sum_j Q_j } ,\quad \left[ T(\beta) , U(\theta) \right] = 0.
\end{align*}
For future reference, we introduce the tensor $u(\theta)$
\begin{equation} \label{eq:Utensor}
\diagram{intro}{6} = \delta_{n_1}^{n_2} \exp\left( i \theta n_1 \right),
\end{equation}
such that the symmetry operation is $U(\theta)=\bigotimes_i u_i(\theta)$.
\par The Mermin-Wagner theorem now dictates that this $\mathrm{U}(1)$ symmetry cannot be broken at any finite temperature. On the level of the transfer matrix this implies that the leading eigenvector (fixed point) necessarily is invariant under the $\mathrm{U}(1)$ transformation. For the MPS approximation of the fixed point, this implies that
\begin{multline*}
\diagram{fixedpoint}{6} \\ = \diagram{fixedpoint}{1} 
\end{multline*}
for any $\theta$. The fundamental theorem of MPS now dicates that we can associate this symmetry of the MPS to a symmetry property of the local tensor $A$ \cite{PerezGarcia2006}. Specifically, we have that the virtual legs of the MPS tensor transform themselves according to representations of the $\mathrm{U}(1)$ symmetry,
\begin{equation} \label{eq:virtualrep}
\diagram{fixedpoint}{7} = \diagram{fixedpoint}{8}.
\end{equation}
In general, these representations on the virtual level can be projective. Imposing that the MPS is invariant under group transformations implies that the tensor $A$ has a certain block structure, where each block can be labeled by a choice of representations on each leg of the tensor.

\par Although the characteristic physics of the XY model is determined by the $\mathrm{U}(1)$ symmetry, we can also take into account the charge-conjugation symmetry $C$ that flips the sign of the charges in the transfer matrix. The total symmetry group that we consider is the (non-abelian) group $\mathrm{U}(1)\rtimes C$, which corresponds to the full $\mathrm{O}(2)$ symmetry of the XY model. There are two types of irreducible representations (irreps) of this $\mathrm{O}(2)$ symmetry. First, there are the irreps corresponding to integer charges $n=0,1,2,\dots$; for $n=0$ there are two one-dimensional irreps, whereas for $n>0$ they are all two-dimensional. Second, there are half-integer charges $n=\frac{1}{2},\frac{3}{2},\dots$, corresponding to projective representations; these irreps are all two-dimensional. 
\par Since the transfer matrix only contains legs with integer representations, the `physical legs' of the MPS all transform according to integer representations as well. This, however, still allows for the freedom on the virtual level of the MPS to impose either integer representations or half-integer representations, giving rise to two different classes of $\mathrm{O}(2)$ invariant MPS. In gapped phases, the occurence of either set of representations is known to characterize a trivial (integer) or symmetry-protected topological (half-integer) phase \cite{Pollmann2010, Chen2011}. In order to illustrate these different classes of MPS and to see which one is realized for the fixed point of the XY transfer matrix, we have performed simulations (i) without imposing symmetries on the legs, (ii) with integer irreps on the legs, and (iii) with half-integer irreps on the legs. In Table \ref{tab:charges} we list the corresponding values for the transfer-matrix leading eigenvalue $\lambda$ at a point in the critical phase.
\par A first observation is that we find a higher value for $\lambda$ when no symmetries are imposed at significantly lower bond dimension. This shows that a variational MPS favours the breaking of the $\mathrm{U}(1)$ symmetry. As we have discussed in the introduction, this artificial symmetry breaking is also observed in other critical systems with a $\mathrm{U}(1)$ symmetry \cite{Draxler2013}. The explanation for this effect can be sought in the fact that MPS necessarily induces a finite correlation length. Indeed, simulating a critical system with MPS with a finite bond dimension can be thought of as slightly perturbing the system such that a gap is opened and a finite correlation length is introduced. From the perspective of an effective field theory describing the critical system, the MPS adds a relevant perturbation that opens up a gap in this field theory. In the $\mathrm{U}(1)$ phase, however, only symmetry-breaking terms are relevant, such that we expect that the MPS approximation induces an artificial symmetry breaking. This observation is confirmed when we compute the $\mathrm{U}(1)$ order parameter in the next section.
\par Secondly, upon imposing the $\mathrm{U}(1)$ or $\mathrm{O}(2)$ symmetry explicitly, we observe that the eigenvalue reaches a similar value at a comparable bond dimension in the two (normal and projective, resp.) sectors. If we again interpret the MPS approximation as introducing a gap, this result suggests that the critical $\mathrm{U}(1)$ phase can be perturbed into both an SPT phase and a regular phase. As we will see in Sec.~\ref{sec:gapped}, in the gapped phase it can be determined without ambiguity what irreps should be chosen on the MPS virtual legs.

\begin{table}
\begin{tabular}{ | p{3cm} | p{1cm} | p{2cm} | }
\hline
& $D$  & $\lambda$ \\
\hline
no symmetries & 19 &        2.5869206 \\
integer charges & 45 &      2.5869172 \\
half-integer charges & 46 & 2.5869184 \\
\hline
\end{tabular}
\caption{The leading eigenvalue (per site) of the XY transfer matrix in the critical phase ($T=0.8$) as obtained by imposing an MPS approximation for the fixed point, where we have imposed different symmetry properties on the virtual legs of the MPS. }
\label{tab:charges}
\end{table}

\subsection{Computing local observables}
\label{sec:observables}

Local observables such as the internal energy and the magnetization can be represented in the tensor-network language as follows. A generic one-angle observable at site $j$ 
\begin{equation*}
\braket{h(\theta_j)} = \frac{1}{Z} \prod_i\left(\int\frac{\d\theta_i}{2\pi}\right) \e^{-\beta E(\{\theta_i\})}  h(\theta_j),
\end{equation*}
can, under the duality transform that we introduced above, be represented diagrammatically as
\begin{equation*}
\braket{h(\theta_i)} = \frac{1}{Z} \left( \diagram{observ}{1} \right) 
\end{equation*}
with
\begin{multline*}
\diagram{observ}{2} = \left( \prod_{i=1}^4I_{n_i}(\beta) \right)^{1/2}  \\ \times \int \frac{\d\theta}{2\pi} \; \e^{\beta h \cos\theta} \e^{i\theta(n_{1}+n_{2}-n_{3}-n_{4})} h(\theta).
\end{multline*}
Using the MPS representation of the transfer-matrix fixed point, we can simplify this to
\begin{equation*}
\braket{h(\theta)} = \frac{\left(\diagram{observ}{3}\right)}{\left(\diagram{observ}{4}\right)}\;.
\end{equation*}
The contractions of these infinite channels are evaluated by finding the leading eigenvectors of the channel operators. 
\par
\begin{figure} \centering
\subfigure[]{\includegraphics[width=0.94\columnwidth]{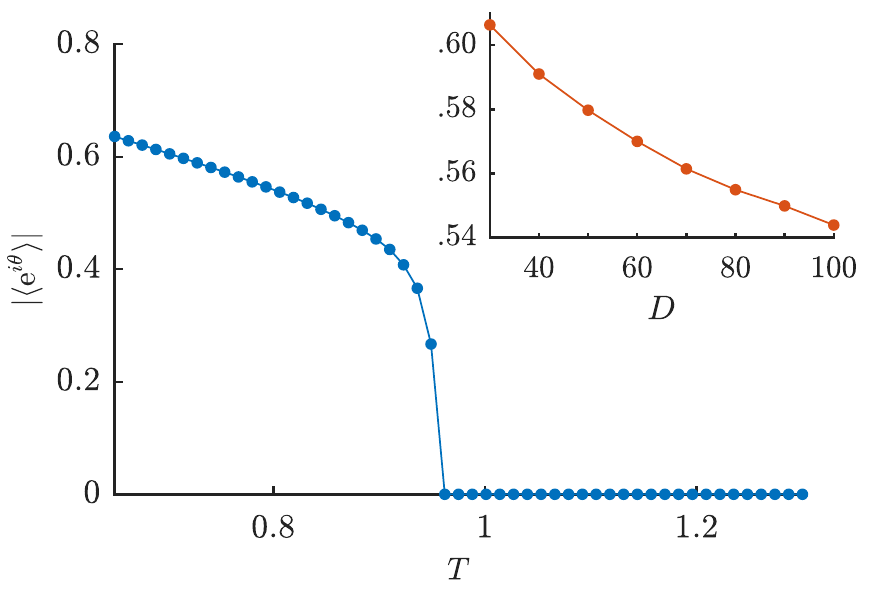}}\\
\subfigure[]{\includegraphics[width=0.94\columnwidth]{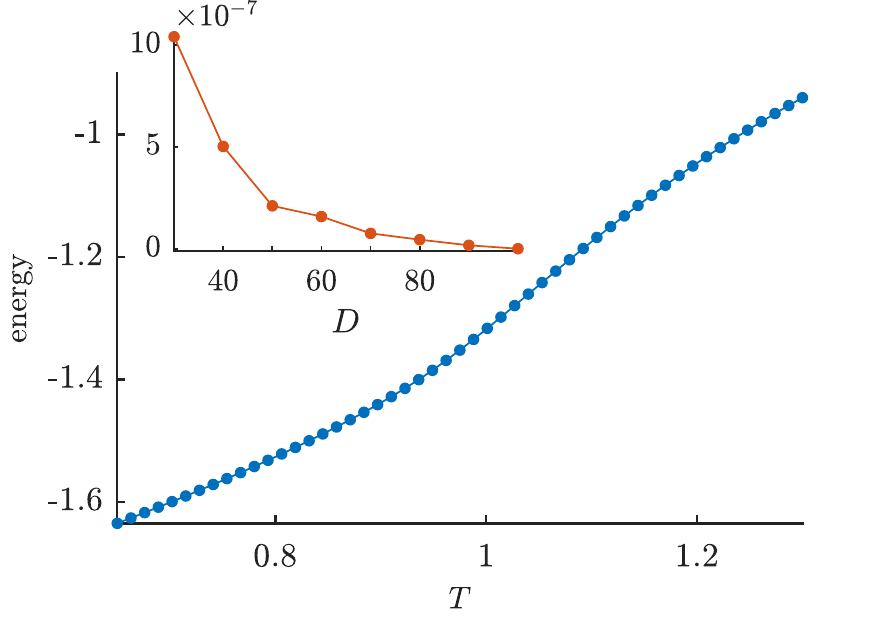}} \\
\subfigure[]{\includegraphics[width=0.94\columnwidth]{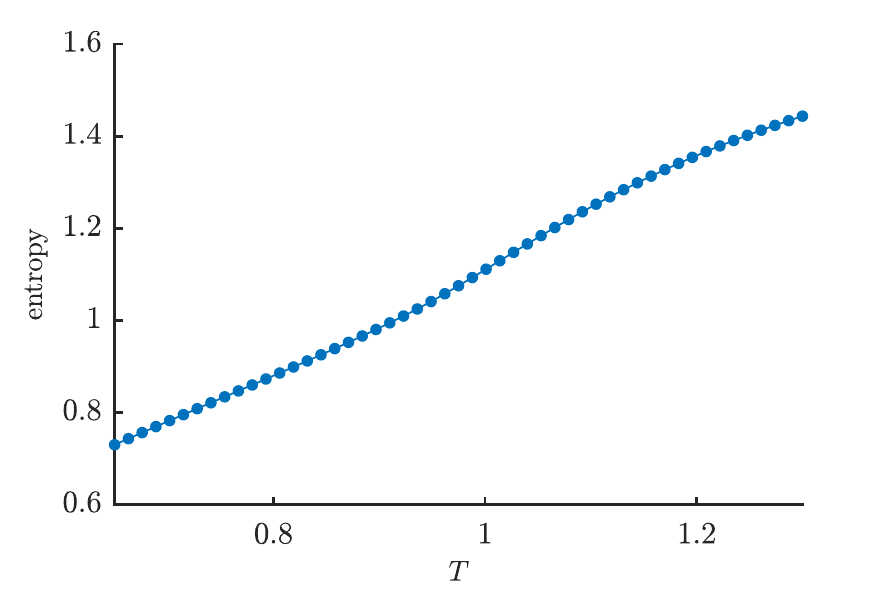}} 
\caption{Observables from MPS simulations with bond dimension $D=30$ (no virtual symmetries). In (a) we plot the order parameter, showing significant symmetry breaking in the critical region; the inset shows that this value (for $T=0.7$) of the order parameter decreases very slowly with the bond dimension. In (b) we plot the internal energy, showing no sign of a phase transition; here, the inset shows good convergence even for small bond dimensions ($T=0.7$). In (c) we plot the entropy per site, which is easily evaluated from the free and internal energy; we have renormalized the entropy as $s(\beta) \to s(\beta)-s(\infty)$, with $s(\infty)=\log(\frac{1}{2\pi})$, such that the zero-temperature limit yields a zero entropy and the entropy is positive everywhere.}
\label{fig:observables}
\end{figure}

 Similarly, a generic nearest-neighbour two-angle observable at sites $j$ and $k$ is given by
\begin{equation*}
\braket{h(\theta_j,\theta_k)} = \frac{1}{Z} \left( \diagram{observ}{5} \right),
\end{equation*}
with
\begin{align*}
& \diagram{observ}{6} = \left( \prod_{i=1}^6 I_{n_i}(\beta) \right)^{1/2} \\ 
& \qquad \times \int \frac{\d\theta_j}{2\pi} \int \frac{\d\theta_k}{2\pi} h(\theta_j,\theta_k) \e^{\beta h (\cos\theta_j+\cos\theta_k)} \\
& \qquad  \times \sum_m \left( I_m(\beta) \e^{i\theta_j(n_{1}+m-n_{5}-n_{6})} \right. \\ & \qquad\hspace{3cm} \left.  \e^{i\theta_k(n_2+n_3-n_4-m)} \right).
\end{align*}
Using the MPS fixed points that we have optimized earlier for computing the free energy -- i.e. without using symmetries on the virtual level -- we now compute the internal energy $e$, the entropy $s$ and the order parameter $o$
\begin{align*}
e &= -2 \braket{\cos(\theta_i-\theta_j)} \\
s &= \beta(e-f) \\ 
o &= \left| \braket{\e^{i\theta_i}} \right|,
\end{align*}
as a function of temperature, and plot the results in Fig.~\ref{fig:observables}. Again, the energy and entropy show no sign of the phase transition. Also, we observe that the value for the energy has already converged up to an error of $\epsilon=10^{-6}$ at a bond dimension of $D=30$; since the entropy is computed from the free energy and the internal energy, it has the same accuracy. As anticipated in the previous section, the order parameter shows a very large value in the critical region, which decreases very slowly as the bond dimension increases. This shows that the MPS breaks the continuous $\mathrm{U}(1)$ symmetry significantly in the critical phase, whereas in the gapped phase the symmetry is restored. The fact that the order parameter decays very slowly with increasing bond dimension shows that this is an essential property of MPS approximations for $\mathrm{U}(1)$ phases. On the other hand, from the convergence of the free and internal energy, we see that this does not prohibit an accurate evaluation of the system's physical properties. Note that the order parameter is identically zero if we would impose the $\mathrm{U}(1)$ or $\mathrm{O}(2)$ symmetry on the virtual level of the MPS, but, as we have seen in Tab.~\ref{tab:charges}, at a large variational cost in the free energy.

\section{The spin stiffness}

The phase transition in the XY model can, in the absence of a local order parameter, be characterized by the so-called spin stiffness. This quantity is defined as the response to a twist field $\vec{v}$, which rotates the angles as
\begin{equation*}
\theta_i \rightarrow \theta_i + \vec{v} \cdot \vec{n}_i.
\end{equation*} 
If we take the twist field along the $y$ axis, this modifies the classical hamiltonian (without magnetic field) to
\begin{equation*}
H_v = -\sum_{\braket{ij}_x} \cos(\theta_i-\theta_j ) - \sum_{\braket{ij}_y} \cos(\theta_i-\theta_j+v) .
\end{equation*}
On the level of the partition function, this introduces an extra phase factor on the vertical links,
\begin{multline*}
Z_v = \prod_{l\in\mathcal{L}_y}  \left( \sum_{n_l=-N}^N I_{n_l} (\beta) \e^{in_lv} \right) \\ \times \prod_{l\in\mathcal{L}_x} \left( \sum_{n_l=-N}^N I_{n_l} (\beta) \right)  \prod_{s} F_{n_{s,1},n_{s,2}}^{n_{s,3},n_{s,4}}.
\end{multline*}
Incorporating these extra phase factors, we can represent the partition function as
\begin{equation*}
Z_v = \diagram{rhoS}{1},
\end{equation*}
where the tensor $u_v$ was defined in Eq.~\eqref{eq:Utensor}. We find for the transfer matrix after the twist
\begin{align*}
T_v(\beta) &= \diagram{rhoS}{2} \\
&= T(\beta) U(v) = U(v) T(\beta),
\end{align*}
because $U(v)$ commutes with the transfer matrix. This implies that all eigenvectors of the transfer matrix remain unchanged after a twist and the eigenvalues are multiplied by a phase. In particular, the fixed point remains unchanged and, as the fixed point lives in the charge-zero sector, so does the leading eigenvalue. This, in turn, implies that the free energy density is a constant when applying a twist
\begin{equation*}
f(v) = f(0),
\end{equation*}
and that the spin stiffness, defined as
\begin{align*}
\rho = \left. \frac{\partial^2 f}{\partial v^2} \right|_{v=0},
\end{align*}
is identically zero for all temperatures. 
\par This is a surprising result, because $\rho$ is supposed to jump discontinuously at the transition. The reason for this discrepancy lies in the fact that we work directly in the thermodynamic limit. Indeed, the spin stiffness is typically defined in a system that is finite and has periodic boundary conditions in the direction of the twist. In that setting, applying the twist field is equivalent to imposing twisted boundary conditions -- this can be observed from the above representation of the partition function, where the operator $U(v)$ can be pulled through the lattice -- and is, therefore, a finite-size property. It is only after it has been defined on this periodic system, that the infinite-size limit can be taken. The above definition of the spin stiffness in terms of the infinite-size transfer matrix, on the other hand, assumes open boundary conditions. If the transfer matrix has no gap in the thermodynamic limit, both definitions are not equivalent.

\begin{figure*} \begin{center}
\includegraphics[width=1.5\columnwidth]{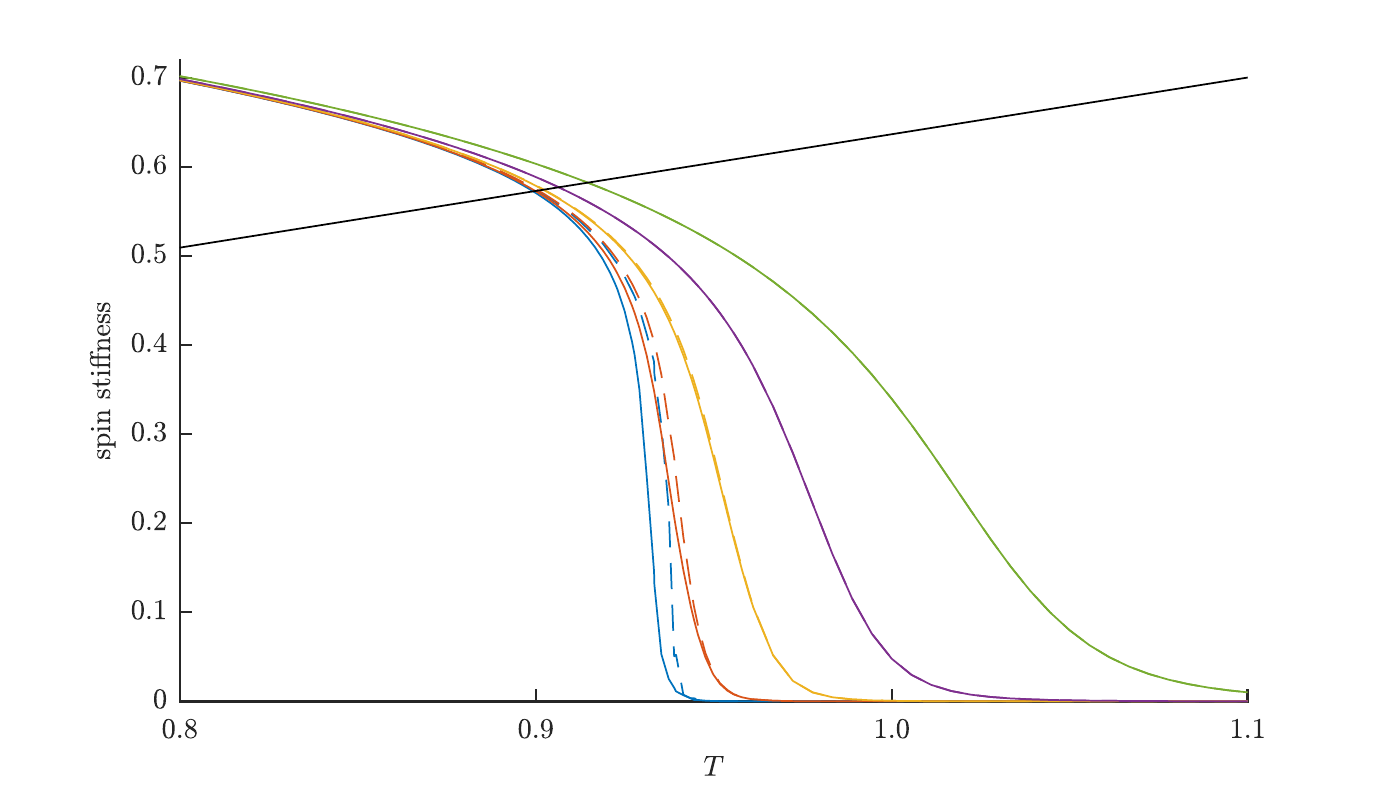}
	\caption{The spin stiffness as a function of temperature for a set of different values of the magnetic field: $h=10^{-3}$ (green), $h=10^{-4}$ (purple), $h=10^{-5}$ (yellow), $h=10^{-6}$ (red), $h=10^{-7}$ (blue). The full lines were computed with a bond dimension $D=150$, the dashed lines are $D=90$. We have also plotted the straight line $\frac{2T}{\pi}$, which is known to intersect the curve for the critical temperature [Eq.~\eqref{eq:stiff}]; for $h=10^{-7}$ and $D=150$ we find an intersection at $T=0.899$.}
    \label{fig:stiffness_curve}
\end{center} \end{figure*}

\par We can work around this by computing the spin stiffness as a two-point function. First we note that we can rewrite the expression for the spin stiffness as 
\begin{align*}
\rho = - \lim_{N\to\infty} \frac{1}{N \beta Z(0)} \left. \frac{\partial^2 Z_v}{\partial v^2} \right|_{v=0},
\end{align*}
where the twisted partition function can be written as
\begin{equation*}
Z_v = \diagram{rhoS}{3},
\end{equation*}
with the twisted tensor 
\begin{equation*}
\diagram{rhoS}{4} = \left( \prod_{i=1}^4I_{n_i}(\beta) \right)^{1/2}  \e^{ in_3v} F_{n_1,n_2}^{n_3,n_4}.
\end{equation*}
In the tensor-network representation of $Z_v$ we can easily differentiate with respect to $v$. Indeed, the first derivative is given by
\begin{equation*}
\frac{1}{N} \left. \frac{\partial Z_v}{\partial v} \right|_{v=0} = \diagram{rhoS}{5},
\end{equation*}
with the tensor $R=\left.\frac{\partial O_v}{\partial v}\right|_{v=0}$, or
\begin{equation*}
\diagram{rhoS}{6} =  in_3 \left( \prod_{i=1}^4I_{n_i}(\beta) \right)^{1/2} F_{n_1,n_2}^{n_3,n_4}.
\end{equation*}
For the second derivative we have to differentiate two different tensors, and twice the same tensor. The result is given by (the site $i$ is arbitrary)
\begin{multline} \label{eq:sf}
\frac{1}{N} \left. \frac{\partial^2 Z_v}{\partial v^2} \right|_{v=0} =  \diagram{rhoS}{8} \\ + \sum_{j \neq i} \diagram{rhoS}{7} \;,
\end{multline}
where we have introduced the tensor $S=\left.\frac{\partial^2 O_v}{\partial v^2}\right|_{v=0}$, or
\begin{equation*}
\diagram{rhoS}{9} = \left( in_3 \right)^2 \\ \times \left( \prod_{i=1}^4I_{n_i}(\beta) \right)^{1/2} F_{n_1,n_2}^{n_3,n_4}.
\end{equation*}
This reduces the spin stiffness to a summation of two-point functions, and therefore has the form of a structure factor. We should note that bringing the factor $in_3$ down in the tensor is equivalent to introducing a factor $\sin(\theta_i-\theta_j)$ in the partition function because of the identities
\begin{align*}
& \sum_{n=-\infty}^\infty (in) I_n(x) \e^{in\theta} = \e^{x\cos\theta} \sin\theta   \\
& \sum_{n=-\infty}^\infty (in)^2 I_n(x) \e^{in\theta} = \e^{x\cos\theta} (\sin^2\theta -\cos\theta) .
\end{align*}
The spin stiffness can therefore be brought into the form
\begin{equation*}
\rho = - \frac{1}{N \beta} \left( \braket{S_y^2} - \braket{\cos(\theta_i-\theta_j)} \right)
\end{equation*}
with
\begin{equation*}
S_y = \sum_{\braket{ij}_y}\sin(\theta_i-\theta_j).
\end{equation*}
In this form, it can be evaluated in Monte-Carlo simulations on a system with periodic boundary conditions without explicitly applying a twist \cite{Sandvik2010, Hsieh2013}. 
\par Again, the evaluation of this two-point function is identically zero on an infinite system with open boundary conditions in the presence of an unbroken $\mathrm{U}(1)$ symmetry. Indeed, if we represent the infinite upper plane of the above expression by the fixed point of the transfer matrix, we observe that
\begin{multline*}
\sum_j \diagram{rhoS}{10} \\ = i \sum_j \diagram{rhoS}{11},
\end{multline*}
which vanishes because the fixed point is $\mathrm{U}(1)$ symmetric. This is, of course, consistent with the vanishing of the spin stiffness as defined above. 
\par For evaluating the spin stiffness, we can, however, introduce a magnetic field in the hamiltonian, which breaks the $\mathrm{U}(1)$ symmetry of the model and induces a gap. In the presence of this extra field, the spin stiffness as defined by the above structure factor [Eq.~\eqref{eq:sf}] can be evaluated in an infinite system with open boundary conditions. Similarly to taking the infinite-size limit for a periodic system, we can then take the limit $h\rightarrow0$ to obtain the result in the zero-field case. In Fig.~\ref{fig:stiffness_curve} the results for the spin stiffness (as defined from the structure factor in Eq.~\eqref{eq:sf}) as a function of temperature are given for a few values of the magnetic field. These results were obtained by using the channel-environment construction of Ref.~\onlinecite{Vanderstraeten2015b} for evaluating structure factors in two-dimensional tensor networks. The figure shows that the drop in the spin stiffness becomes sharper as the magnetic field is decreased, but for very small values of $h$ the effects of a finite bond dimension become more pronounced.

\section{Luttinger liquid mapping}
\label{sec:luttinger}

In the previous section we have computed the spin stiffness by explicitly breaking the $\mathrm{U}(1)$ symmetry via an external magnetic field. In this section, we show that we can use the Luttinger-liquid formalism \cite{Giamarchi2004} to compute the same quantity without an explicit breaking of the symmetry, directly in the thermodynamic limit.
\par Early on, it was realized \cite{Jose1977} that the proper effective field theory for the partition function is given by the sine-Gordon model \cite{Benfatto2013}, which is described by the one-dimensional quantum hamiltonian
\begin{multline*}
H_{\text{SG}} =  \frac{1}{2\pi} \int dx \left( uK \left( \nabla \theta(x) \right)^2 + \frac{u}{K} \left(\nabla\phi(x)\right)^2 \right) \\ + g \int dx \cos(2\phi(x)).
\end{multline*}
The first line is the hamiltonian of the Luttinger-liquid field theory \cite{Giamarchi2004} describing the critical spin-wave excitations, and the second line adds vortices to the picture. The microscopic $\mathrm{U}(1)$ symmetry of the model is reflected in the sine-Gordon field theory by the generator
\begin{equation*}
 - \frac{1}{\pi} \int d x \nabla\phi(x).
\end{equation*}
Since vortices are irrelevant in the critical phase, we expect that under an RG transformation the vortices will drop from the sine-Gordon hamiltonian and the parameters $u$ and $K$ will be renormalized. In other words, we expect that the low-energy properties of the XY model will be described by an effective Luttinger-liquid field theory 
\begin{equation*}
H_{\text{LL}} =  \frac{1}{2\pi} \int dx \left( \tilde{u}\tilde{K} \left( \nabla \theta(x) \right)^2 + \frac{\tilde{u}}{\tilde{K}} \left(\nabla\phi(x)\right)^2 \right),
\end{equation*}
with the effective parameters $\tilde{u}$ and $\tilde{K}$ determined by the inverse temperature $\beta$. As soon as the effective Luttinger parameter reaches $\tilde{K}=2$, however, vortices become relevant, leading to a gapped phase where the $\phi$ field is locked in the minima of the cosine term; the elementary excitations are kinks and anti-kinks between the different ground states.

\begin{figure} \centering
\subfigure[]{\includegraphics[width=0.93\columnwidth]{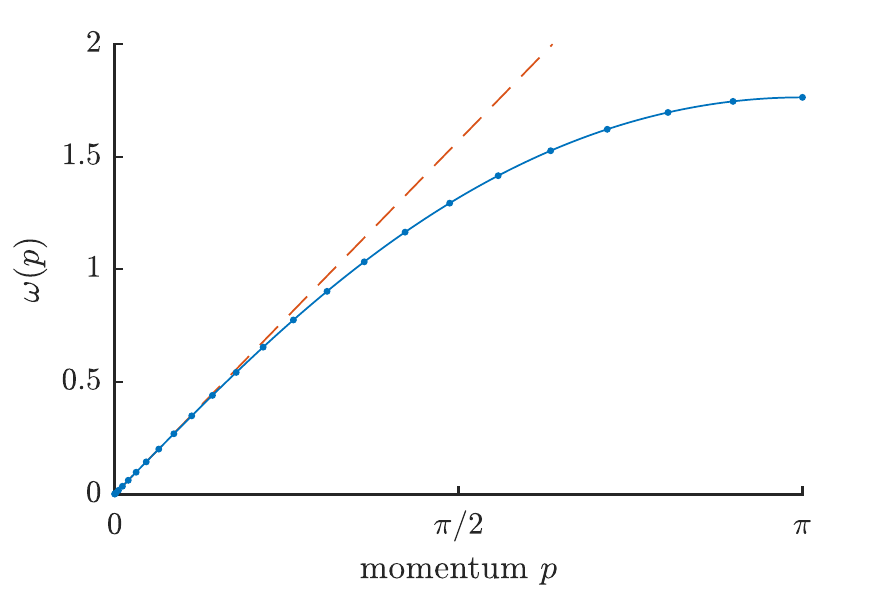}}\\
\subfigure[]{\includegraphics[width=0.93\columnwidth]{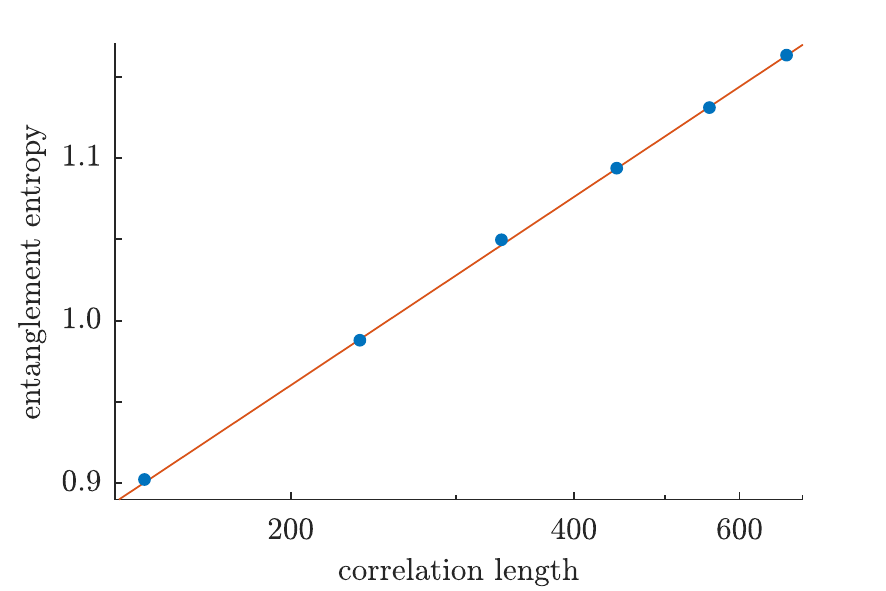}}\\
\subfigure[]{\includegraphics[width=0.93\columnwidth]{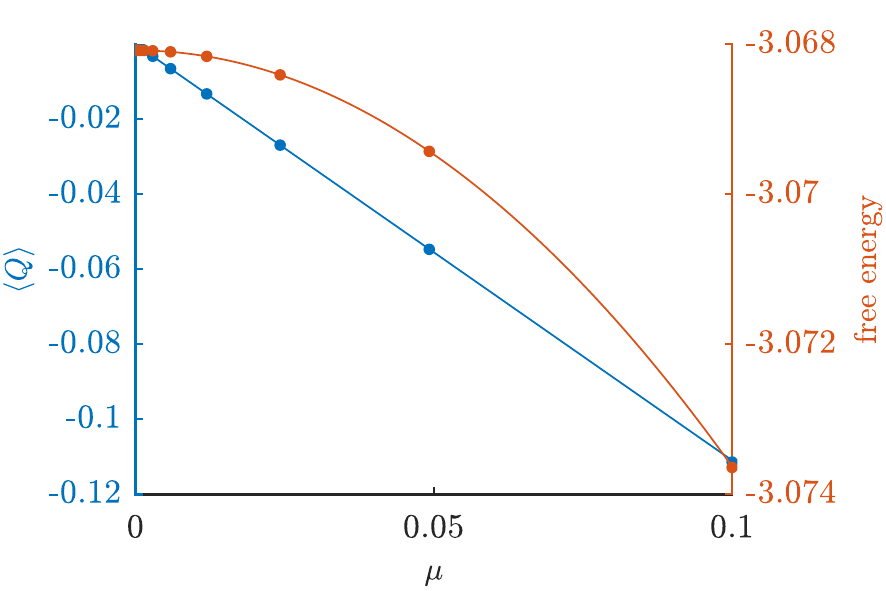}} 
\caption{The XY model at $T=0.7$ as a Luttinger liquid. In (a) we have plotted the dispersion relation of the transfer matrix as defined by Eq.~\eqref{eq:minuslog}; the dashed line is a straight line with a slope equal to one, showing that we find a Luttinger velocity exactly equal to one. In (b) we have plotted the correlation length versus the entanglement entropy obtained in MPS approximations for the fixed point with increasing bond dimension ($D=50:25:175$); the fit is made from the three last points according to the form Eq.~\eqref{eq:entanglement}, yielding a value for the central charge $c=1.003$. In (c) we have plotted the response in the free energy (left) and the expectation value for $Q_i$ with a quadratic and linear fit, resp. The same value $\kappa\approx1.10953$ is found from both fits.}
\label{fig:luttinger}
\end{figure}

\par We can lift this notion of an effective field theory to the level of a transfer matrix, where we formally introduce a Luttinger-liquid transfer matrix
\begin{equation*}
T_{\text{LL}} = \exp \left( - H_{\text{LL}} \right).
\end{equation*}
The idea is that this captures the low-energy behavior of the XY transfer matrix, where the parameters $\tilde{u}$ and $\tilde{K}$ depend on the inverse temperature $\beta$. We confirm the low-energy correspondence of the XY transfer matrix $T(\beta)$ with this effective Luttinger-liquid form in three different ways.
\par The first correspondence can be found by computing the low-lying spectrum of the XY transfer matrix, using the MPS quasiparticle ansatz for the low-lying excited states \cite{Haegeman2016b, Vanderstraeten2019}. We can label the excitations with a momentum $p$, yielding a dispersion relation $\omega(p)$ as the logarithm of its eigenvalues,
\begin{equation} \label{eq:minuslog}
\omega(p) = -  \log \left( \frac{\bra{\phi(p)} T(\beta) \ket{\phi(p)}}{\braket{\phi(p)|\phi(p)}} \right),
\end{equation}
and observe that we find a gapless spectrum with a linear dispersion relation $\omega(p) = \tilde{u} |p|$. Even stronger, we find the effective velocity $\tilde{u}=1$ for all values of the temperature within the critical phase (see Fig.~\ref{fig:luttinger}(a)). It is, of course, expected that the effective field theory is isotropic, but this serves as an excellent test of our transfer-matrix approach.
\par A second test consists of measuring the central charge, which can be determined in an MPS simulation by comparing the scaling of the entanglement entropy and the correlation length as a function of the bond dimension. It is known that, for a system that is described by a conformal field theory the scaling is determined by the central charge $c$ as \cite{Pollmann2009, Pirvu2012}
\begin{equation} \label{eq:entanglement}
S_D \propto \frac{c}{6} \log( \xi_D ).
\end{equation}
In Fig.~\ref{fig:luttinger}(b) we show clear evidence for a central charge $c=1$, which is precisely the value for a Luttinger liquid.
\par An estimate for the effective Luttinger parameter $\tilde{K}$ is obtained by computing the response to a chemical potential. It is easily seen \cite{Giamarchi2004} that adding the generator of the $\mathrm{U}(1)$ symmetry,
\begin{equation*}
T_{\text{LL}} \to \exp \left( \frac{\mu}{\pi} \int dx \nabla \phi(x) \right) T_{\text{LL}}
\end{equation*}
yields a compressilbity that is equal to
\begin{equation*}
\kappa = - \frac{1}{\pi}  \frac{d}{d\mu}  \braket{\nabla\phi}_\mu = \frac{\tilde{K}}{\tilde{u}\pi}.
\end{equation*}
We have an explicit microscopic form for the generator of the $\mathrm{U}(1)$, so we can explicity implement a chemical potential on the level of the microscopic transfer matrix as
\begin{equation*}
T_\mu(\beta) = \exp\left( -\frac{\mu}{2} Q \right) T(\beta)  \exp\left( -\frac{\mu}{2} Q \right).
\end{equation*}
This shift in the transfer matrix does not affect its eigenvectors, but reshuffles the eigenvalues. Therefore the fixed point $\ket{\psi_\mu}$ will change as a function of $\mu$, and will give rise to a finite expectation value of the generator. The associated compressbility
\begin{equation*}
\kappa = \left. \frac{d \bra{\psi_\mu} Q_i \ket{\psi_\mu} }{d \mu} \right|_{\mu=0}
\end{equation*}
will therefore give a finite value. This, in turn, yields a direct estimate of the effective Luttinger parameter $\tilde{K}$. Since $Q$ commutes with the transfer matrix, the partition function will only be affected in second order, and the second derivative yields the same value for the compressibility,
\begin{equation*}
\kappa = - \left. \frac{d^2 \log \lambda_\mu(\beta)}{d\mu^2}  \right|_{\mu=0},
\end{equation*}
where $\lambda_\mu(\beta)$ is the usual eigenvalue per site of the transfer matrix
\begin{equation*}
\lambda_\mu(\beta) = \lim_{N_x\to\infty} \bra{\psi_\mu} T_\mu(\beta) \ket{\psi_\mu} ^{1/N_x}.
\end{equation*}
We can compute this compressibility straightforwardly using the uniform MPS framework. Indeed, as explained above, the eigenvalue $\lambda_\mu(\beta)$ is the quantity that is variationally optimized in an MPS fixed-point simulation, whereas the charge density $\braket{Q_i}$ can be easily computed as an expectation value. In Fig.~\ref{fig:luttinger} it is shown that both quantities yield a consistent numerical value for the compressibility.
\par We should note, however, that the transfer matrix $T_\mu(\beta)$ is equivalent to the one of the twisted XY model but with an imaginary value for the twist field
\begin{equation*}
T_\mu(\beta) = T(\beta) U(i\mu).
\end{equation*}
Therefore, the effective Luttinger parameter that we have defined here is related to the spin stiffness from before,
\begin{equation*}
\tilde{K} = \pi\beta\rho.
\end{equation*}
This correspondence, which can be readily seen from the mapping of the XY model to the sine-Gordon field theory \cite{Jose1977, Benfatto2013}, yields the famous \cite{Nelson1977} value for the spin stiffness at the critical temperature $\rho=2T_c/\pi$. 
\par In the previous section, we had anticipated that computing the spin stiffness without introducing a symmetry-breaking term would not be possible in the thermodynamic limit directly. The reason that we are here able to compute the spin stiffness without breaking the $\mathrm{U}(1)$ symmetry consists in the fact that we have expressed it as a thermodynamic quantity (the compressibility) for which the extensivity properties of the uniform MPS simulations are ideally suited. This thermodyamic quantity, however, is necessarily formulated on the level of the transfer matrix after the duality transformation, since an imaginary twist does not translate to a realistic modification of the classical XY hamiltonian.
\par
\begin{figure} \centering
\subfigure[]{\includegraphics[width=0.95\columnwidth]{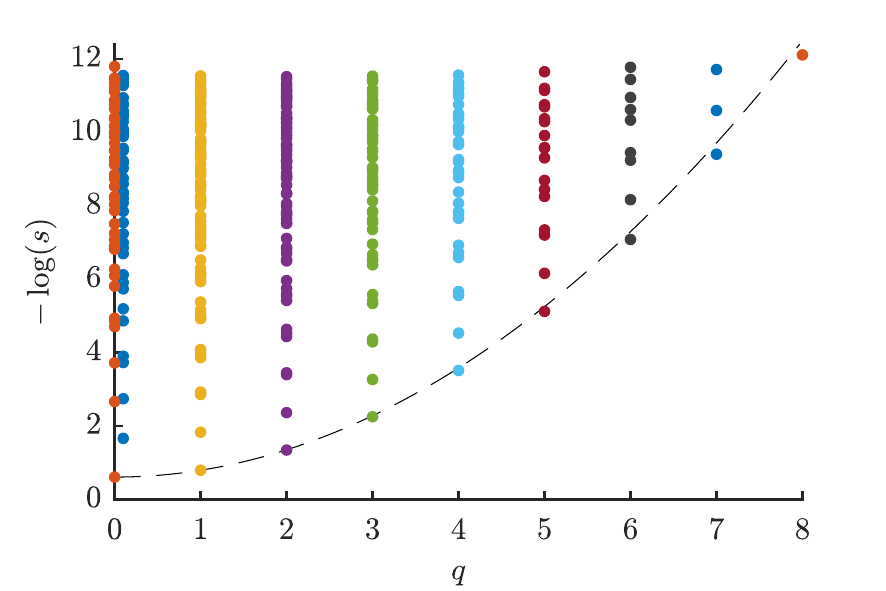}}\\
\subfigure[]{\includegraphics[width=0.95\columnwidth]{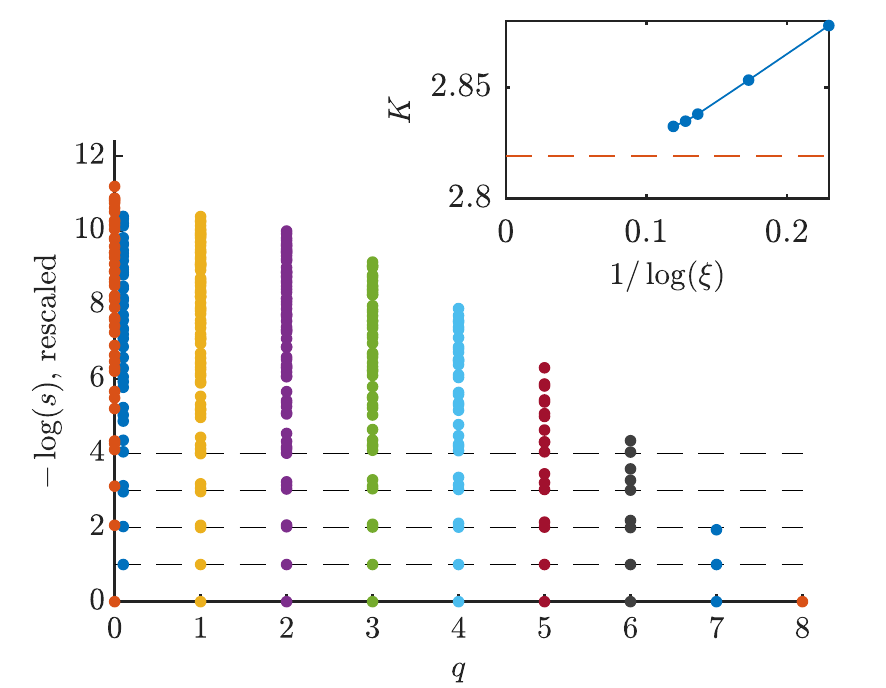}}
\caption{Entanglement spectrum of the fixed-point MPS of the XY transfer matrix at $T=0.8$. We have imposed the full $\mathrm{O}(2)$ symmetry on the MPS tensor, which implies that the entanglement spectrum is labeled by the irreps on the virtual bonds: two one-dimensional irreps with charge $q=0$ (blue, red), and two-dimensional irreps with charges $q=1,2,\dots$. In (a) we plot the bare entanglement spectrum, and we find that the envelope of the entanglement spectrum follows a quadratic form (striped line). In (b) we have shifted the different sectors such that the lowest value is zero; this produces a nice free-boson boundary CFT spectrum.}
\label{fig:cft}
\end{figure}

 As a final signature of the Luttinger-liquid phase, we investigate the entanglement spectrum of the MPS fixed point. As was observed in Ref.~\onlinecite{Laeuchli2013}, the low-lying part of the entanglement spectrum for a bipartition of the MPS, should resemble the energy spectrum of a boundary conformal field theory (CFT). In Fig.~\ref{fig:cft} we plot the entanglement spectrum of the MPS fixed point, where we have imposed $\mathrm{O}(2)$ symmetry on the MPS tensor such that we can label the spectrum with the appropriate quantum numbers. We observe that the spectrum has a quadratic envelope, and that we obtain an equidistant spectrum after rescaling the different sectors, in perfect correspondence with the spectrum of a free-boson boundary CFT. Moreover, from the rescaling parameter we can deduce an estimate of the Luttinger parameter $\tilde{K}$, which nicely converges to the same value as the one we find using the compressibility.

\section{The gapped phase}
\label{sec:gapped}

The characterization of the gapped phase of the XY transfer matrix using uniform MPS is a lot more straightforward. Indeed, we expect that the fixed point of a gapped transfer matrix can be approximated by an MPS with arbitrary precision. Therefore, we no longer expect that the MPS will spontaneously break the $\mathrm{U}(1)$ or $\mathrm{O}(2)$ symmetry, and we can safely use the fundamental theorem to realize that the virtual legs of the MPS should transform under (projective) representations as well (according to Eq.~\eqref{eq:virtualrep}).
\par
\begin{figure} \centering
\includegraphics[width=0.95\columnwidth]{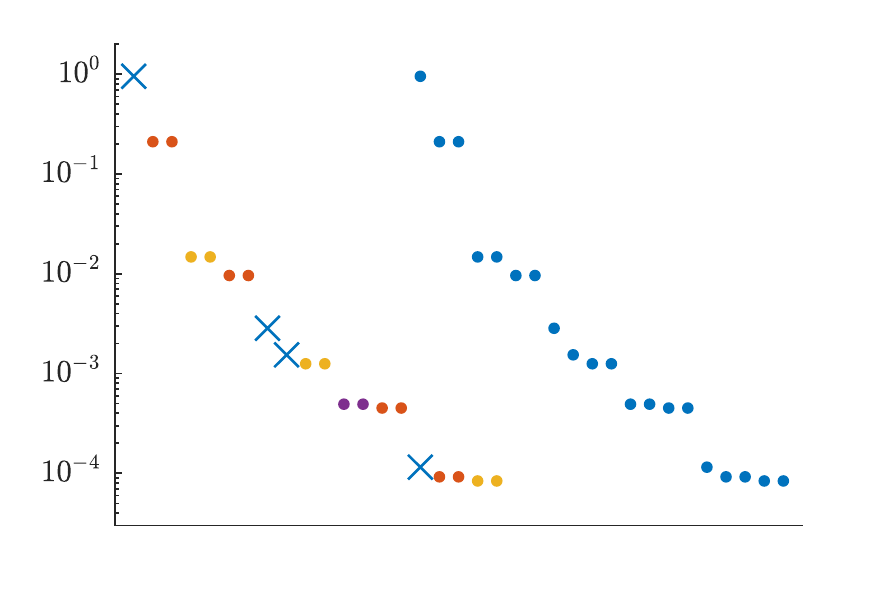}
\caption{The entanglement spectrum of the fixed-point MPS at $T=1.216$, where we impose $\mathrm{O}(2)$ invariance on the virtual legs (left) and without explicit symmetries (right). On the left, we label the different $\mathrm{O}(2)$ representations as follows: $n=0$ (blue crosses), $q=\pm1$ (red), $q\pm2$ (orange) and $q=\pm3$ (purple), whereas on the right we don't have no labeling; we only show the Schmidt values above $10^{-4}$. The perfect correspondence of the entanglement spectra with and without explicit symmetries on the virtual legs shows that the MPS representation for the fixed-point only contains integer representations of $\mathrm{O}(2)$.}
\label{fig:schmidt}
\end{figure}

 As explained in Sec.~\ref{sec:symmetries}, we are ignorant on which representations should be chosen on the virtual legs of the MPS. For that reason, we first plot the entanglement spectrum of a fixed-point MPS around temperature $T=1.2$ without explicit symmetries on the MPS representation (see Fig.~\ref{fig:schmidt}). We find both isolated and twofold degenerate Schmidt values, which point to integer representations on the virtual legs. Indeed, in Fig.~\ref{fig:schmidt} we plot the entanglement spectrum with integer representations on the virtual legs imposed, showing that the isolated values correspond to either of the two one-dimensional irreps with $n=0$ charge sector; the twofold degenerate ones correspond to $n=1,2,\dots$. 

\par
\begin{figure} \centering
\subfigure[]{\includegraphics[width=0.95\columnwidth]{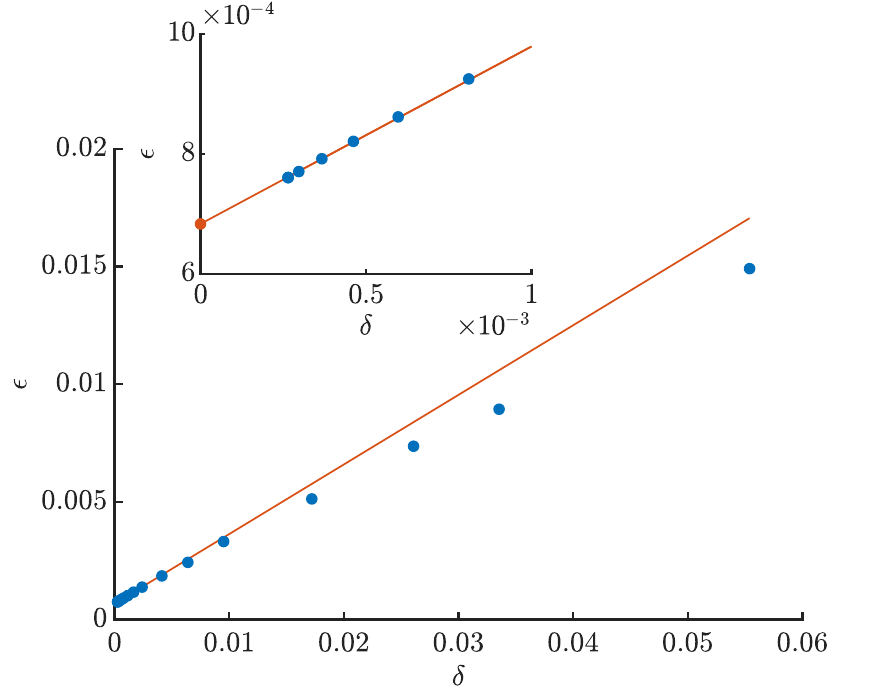}}\\
\subfigure[]{\includegraphics[width=0.95\columnwidth]{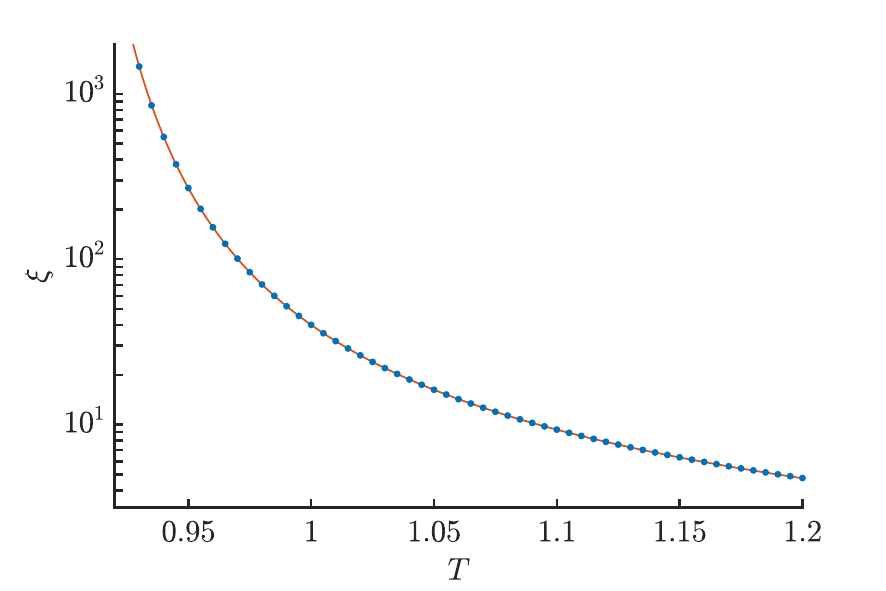}}
\caption{In (a) we plot the extrapolation procedure for $T=0.93$ yielding a value for the correlation length $\xi=1463(4)$; the maximal dimension in each block was set at $D_\mathrm{max}=512$, yielding a total MPS bond dimension of $D=3754$. In (b) we plot the extrapolated correlation lengths as a function of temperature. We fit this to the KT form [Eq.~\eqref{eq:corr2}] (red line), yielding a value for the critical temperature $T_c=0.8930(1)$.}
\label{fig:corrlength}
\end{figure}

 One of the hallmarks of the BKT transition is the exponential divergence of the correlation length when approaching the critical point from the gapped side [Eq.~\eqref{eq:corr}]. Using MPS, we can confirm this behaviour and use this form to obtain an estimate for the critical point. The correlation length is a notoriously hard quantity to converge in MPS simulations, but using the second gap in the transfer matrix (typically denoted as $\delta$) it is possible to extrapolate its value in a reliable way \cite{Rams2018}. In Fig.~\ref{fig:corrlength} we plot this extrapolation procedure for $T=0.93$, yielding an accurate value for a correlation length of more than a thousand sites. In Fig.~\ref{fig:corrlength} we then plot the extrapolated correlation lengths as a function of temperature, and fit this to
\begin{equation} \label{eq:corr2}
\log\xi = \frac{b}{\sqrt{T-T_c}} + c + d\sqrt{T-T_c},
\end{equation}
where the extra terms are added to account for deviations away from the critical point. This fit yields an estimate for the correlation length of $T_c=0.8930(1)$, which agrees well with other numerical results (see Tab.~\ref{tab:results}). We should note, however, that estimates for the critical point can depend strongly on higher-order corrections to the scaling behavior -- for the finite-size extrapolation of the spin stiffness this is clearly the case \cite{Hsieh2013}. We leave the careful incorporation of higher-order contributions to the scaling of the correlation length, and a more accurate estimation of the critical point, for further study.

\begin{table*}
	\begin{tabular}{ | p{6cm} | p{4cm} |}
		\hline
Monte Carlo (1979) \cite{Tobochnik1979} & 0.89 \\ \hline
Monte Carlo (2005) \cite{Hasenbusch2005} & 0.8929(1) \\ \hline
series expansion (2009) \cite{Arisue2009} & 0.89286(8) \\ \hline
Monte Carlo (2012) \cite{Komura2012} & 0.89289(6) \\ \hline
Monte Carlo (2013) \cite{Hsieh2013} & 0.8935(1) \\ \hline
tensor-renormalization group (2014) \cite{Yu2014} & 0.8921(19) \\ \hline
uniform MPS (current work) & 0.8930(1) \\ \hline
	\end{tabular}
	\caption{Numerical estimates for the critical temperature}
	\label{tab:results}
\end{table*}

\par
\begin{figure} \centering
\includegraphics[width=\columnwidth]{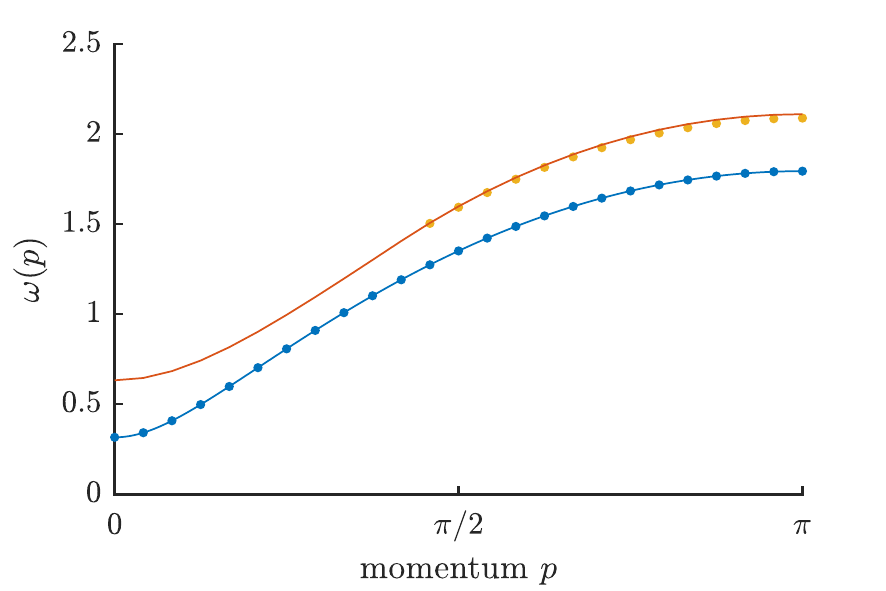}
\caption{The spectrum $\omega(p)$ of the transfer matrix at temperature $T=1.3$. The blue line is elementary excitation branch with charge $q=\pm1$, the red line is the edge of the two-particle continuum. The yellow dots are excitation energies that fall below the continuum edge, signalling a bound state in the $q=0$ sector.}
\label{fig:spectrum}
\end{figure}

 The gapped phase is further characterized by the low-lying spectrum of the transfer matrix, which we define as before in Eq.~\ref{eq:minuslog}. In Fig.~\ref{fig:spectrum} we have plotted the spectrum at a temperature $T=1.3$, showing an isolated two-fold degenerate quasiparticle line; the excited states on this line carry $\mathrm{U}(1)$ quantum numbers $q=\pm1$. Above this elementary one-particle excitation, we find the two-particle continuum and, interestingly, around momentum $p=\pi$ we find a slightly bound state below the continuum with charge $q=0$.

\section{Outlook}
\label{sec:outlook}

In this paper we have investigated the classical two-dimensional XY model using uniform MPS methods. We have shown that an MPS approximation for the fixed point of the XY transfer matrix breaks the $\mathrm{U}(1)$ heavily in the critical phase, which is expected because the MPS always induces a finite correlation length in the system. In a similar vein, a uniform MPS calculation of the spin stiffness in the critical $\mathrm{U}(1)$ phase is always zero. The reason for the latter was sought in the fact that uniform MPS work in the thermodynamic limit directly, whereas the spin stiffness is a quantity that is necessarily defined in a finite periodic system; only for systems with a gap the two definitions intersect.
\par Nonetheless, we showed that uniform MPS are an ideal framework for characterizing the XY model and its phase transition. First of all, despite the fact that an MPS breaks $\mathrm{U}(1)$ symmetry, we can evaluate local observables with very high precision in the critical phase. Secondly, we have shown that the spin stiffness can be evaluated by introducing a small magnetic field $h$, and taking the limit $h\to0$. The mapping to an effective Luttinger-liquid field theory can be made explicit using the MPS framework by computing the central charge, the dispersion relation and the compressibility; the latter, which is the response to a twist with imaginary magnitude, is used to find very accurate values for the effective Luttinger parameter $\tilde{K}$. In addition, we find that the entantglement spectrum of the MPS is in agreement with a boundary CFT spectrum. In the gapped phase, the MPS leaves the (non-abelian) $\mathrm{O}(2)$ symmetry unbroken, which we can use to find accurate values for the correlation length upon approaching the critical point; from fitting the exponential divergence of the correlation length we find $T_c\approx0.8930$, in agreement with other numerical studies.
\par We expect that the current setup can be applied to other two-dimensional classical systems with continuous symmetries. Whereas the standard ferromagnetic Heisenberg model has no phase transition \cite{Shenker1980}, the so-called $\mathrm{RP}^2$ models with a classical hamiltonian $H=-\sum_{\braket{ij}} \left(\vec{s}_i \cdot \vec{s}_j\right)^2$ (with $\vec{s}_i$ a three-dimensional unit vector) potentially hosts $\mathbb{Z}_2$ vortices that drive a phase transition \cite{Solomon1981}. A similar phase transition might be present in the frustrated antiferromagnetic Heisenberg model on the triangular lattice \cite{Kawamura1984, Kawamura2010}. Also, our methods can be readily applied for simulating KT transitions in one-dimensional quantum systems. The relation between these classical topological transitions and one-dimensional SPT phases \cite{DenNijs1989} that we have investigated in this paper might prove very interesting in this context.
\par The results in this paper will prove instrumental in the program of simulating systems with unbroken continuous symmetries with uniform tensor networks and should, in particular, be useful in the study of two-dimensional quantum spin liquids with PEPS. Indeed, the norm of a PEPS can be naturally interpreted as a two-dimensional partition function and the question often poses itself in what phase the corresponding PEPS transfer matrix is. The paradigmatic example here is the resonating valence-bond (RVB) wavefunction on the square lattice, for which the transfer matrix is known to be in a $\mathrm{U}(1)$ phase \cite{Schuch2012, Chen2018a}, and also other symmetric PEPS parametrization for (chiral) spin liquids seem to give rise to critical transfer matrices \cite{Chen2018b}. The relation between the critical properties of the transfer matrix, which are esssentially the properties of a two-dimensional classical system, the symmetries of the PEPS tensors, and the quantum properties of the PEPS wavefunction (physical correlation functions, entanglement spectra, etc.) remains, however, largely unexplored.
\par In a different direction, the two-dimensional partition functions that we have considered here, can be naturally lifted to the quantum level by promoting the charges on the bonds to quantum-mechanical degrees of freedom. Upon doing that, we find quantum-mechanical wavefunctions for $\mathrm{U}(1)$ gauge theories on the lattice. This PEPS construction can be generalized to a whole variational class of states that are ideally suited to study the phase diagram of two-dimensional lattice gauge theories, and, in order to understand the phase transitions, we will need the tools that were explored in this paper.

We acknowledge inspiring discussions with Nick Bultinck, Thierry Giamarchi, Jutho Haegeman, and Masaki Oshikawa. This work was supported by the Flemish Research Foundation, the Austrian Science Fund (ViCoM, FoQuS), and the European Commission (QUTE 647905).

\bibliography{./bibliography}

\end{document}

%% file: preamble.tex
\usepackage{graphicx}
\usepackage{graphics}
\usepackage{amsmath}
\usepackage{amssymb}
\usepackage{amsfonts}
\usepackage{dsfont}
\usepackage{braket}
\usepackage{color}
\usepackage{braket,slashed}
\usepackage[mathscr]{euscript}
\definecolor{darkblue}{rgb}{0, 0, 0.8}
\usepackage[colorlinks=true, breaklinks=true, linkcolor=red, citecolor=blue, urlcolor=blue]{hyperref} 
\usepackage{hyperref}
\usepackage{subfigure}
\usepackage{xfrac}
\usepackage{bm}
\usepackage{kantlipsum}
\usepackage{enumitem}
\usepackage{tikz}
\usepackage{framed}
\usepackage{graphicx}
\usepackage{subfigure}

\allowdisplaybreaks[1]

\newcommand{\code}[1]{\texttt{#1}}

\renewcommand{\d}{\ensuremath{\mathrm{d}}}
\newcommand{\e}{\ensuremath{\mathrm{e}}}

\newcommand{\I}[6]{\ensuremath{I^{#1#2#3}_{#4#5#6}}}

\newcommand{\diagram}[2]{\;\vcenter{\hbox{\includegraphics[scale=0.32,page=#2]{./#1.pdf}}}\;}